\documentclass[12pt]{article}
\usepackage{amsmath,amsfonts}
\usepackage{booktabs}
\usepackage{algorithm}
\usepackage{array}
\usepackage[caption=false,font=normalsize,labelfont=sf,textfont=sf]{subfig}
\usepackage{textcomp}
\usepackage{stfloats}
\usepackage{url}
\usepackage{verbatim}
\usepackage{graphicx}
\usepackage{cite}
\usepackage{algorithm}
\usepackage{algpseudocode}
\usepackage{amsmath, amsfonts}
\usepackage[numbers]{natbib} 
\usepackage[colorlinks=true, linkcolor=blue, citecolor=blue, urlcolor=blue]{hyperref}

\begin{document}

\title{Quantum Radar System Using Born-Feynman path integrals approach}

\author{{Kumar Gautam$^{1^*}$, Akshit Dutta$^{1}$ and Kumar Shubham$^{1}$ 
 \thanks{\\$^{1^*}$ Corresponding Author Email: $\boldsymbol{edumonk@ieee.org}$ \\$^{1}$ Department of Quantum Computing, Quantum Research And Centre of Excellence, Delhi, India.\\Department of Quantum material, Egreen Quanta, India}}}
 
\date{}
\maketitle

\begin{abstract}
The paper relates to a quantum radar deployment by the Born-Feynman path integrals approach based on quantum dots. The radar system comprises a quantum dot-based entangled photon generator, a transmission module, a delay line, a detection module, and a signal processing unit. The quantum dot-based entangled photon generator produces entangled photon pairs via spontaneous parametric down-conversion or stimulated emission. The signal transmission module, equipped with a microwave antenna and beamforming elements, directs the signal photon toward a target. The delay line module synchronizes the retained idler photon with the returning signal photon, preserving quantum coherence. The detection module collects the reflected signal photon and uses a cryogenically cooled superconducting nanowire single photon detector (SNSPD) for detection. Finally, the signal processing unit analyzes the quantum correlation between the scattered and idler photons to enable precise quantum state comparison.

\end{abstract}

\section{Introduction}

Quantum radar is an emerging technology that leverages the principles of quantum mechanics to detect and track objects with greater precision and sensitivity compared to classical radar systems. One promising approach to quantum radar is based on quantum dots (QDs), which are graphene-based nanostructures with discrete energy levels that exhibit quantum mechanical properties \cite{Ref1}-\cite{Ref5}. Quantum radar typically involves the transmission of quantum-entangled photons, 
followed by their interaction with an object and the detection of the reflected (scattered) photons. Quantum radar based on quantum dots technology leverages the principles of quantum entanglement, superposition, and quantum coherence to provide a system with high precision and sensitivity. The key mathematical elements involve the modeling of entangled photon pairs, phase and frequency shifts due to interactions with objects, and signal processing techniques to 
extract distance and velocity information. By exploiting quantum correlations, quantum radar systems can achieve higher resolution and better noise immunity compared to traditional radar systems. Traditional radar technology has been foundational for a wide range of applications, including 
defense and weather monitoring. However, these classical systems face significant challenges concerning resolution, sensitivity, and interference from noise. These issues become particularly problematic in environments where detecting weak signals is necessary or where there are high levels of clutter or noise \cite{Ref6}-\cite{Ref9}. Furthermore, the effectiveness of traditional radar systems is constrained by the existence of stealth technologies and electronic countermeasures designed to evade detection. The invention of the Quantum Radar system based on Quantum Dots technology is presented as a promising solution to address these specific challenges. 

How does quantum radar improve upon traditional radar in terms of detecting capabilities? While operating in difficult environments, quantum radar's detection skills are greatly improved by its many important advantages. Some of the benefits include:

\begin{itemize}
  \item \textbf{Stealth Operation:} Quantum radar can operate in a stealthy manner by using weak signals or quantum states that do not easily reveal their presence to adversaries. This makes it harder to detect or track by enemy radar systems.
  
  \item \textbf{Higher Target Discrimination:} Quantum radar’s ability to utilize entanglement and quantum correlations allows for finer resolution, enabling the radar to distinguish between targets that might otherwise appear as a single, indistinct object in classical radar systems. This improves the radar's ability to identify and track multiple targets.

  \item \textbf{Resistance to Jamming:} Quantum radar can be highly resistant to jamming attempts, particularly in noisy environments, because it can exploit quantum properties like quantum entanglement and squeezing. These techniques help the radar to isolate the actual target signal from the jamming noise, making it more difficult for jammers to interfere with its operation.

  \item \textbf{Enhanced Detection in Noisy Environments:} Quantum radar uses techniques such as quantum illumination, which makes it highly effective in environments with high noise levels, such as urban areas or areas with interference from weather conditions. It can detect targets that are faint and hidden within background noise, which classical radars may struggle to do.

  \item \textbf{Improved Sensitivity to Weak Signals:} Quantum radar can detect very weak signals, even those below the detection threshold of classical radar. This sensitivity makes it ideal for detecting stealth targets or objects with a low radar cross-section (RCS), such as small drones or stealth aircraft.

  \item \textbf{Reduced Power Consumption:} Quantum radar can achieve high levels of sensitivity with lower transmitted power compared to classical radar, leading to reduced energy consumption. This also contributes to its stealthy operation and extended operational range.

  \item \textbf{Better Resolution in Cluttered Environments:} Due to quantum-enhanced sensing, quantum radar can resolve small details of targets, even in cluttered or highly congested environments. This means that it can identify the shape, motion, and other characteristics of objects more effectively than classical radar in such environments.
\end{itemize}

Quantum mechanics will be used to retrieve the data from a quantum radar system. The principles of \textbf{quantum measurement}, \textbf{quantum entanglement}, and \textbf{quantum hypothesis testing} form the basis of data extraction in a quantum radar system. The system stores its entangled partner (\textbf{idler}) and sends out one photon from each entangled pair (\textbf{signal}) in the direction of a prospective target. It is possible to monitor both the signal photon and the idler photon when the former bounces off an object and then returns. The radar assesses the \textbf{quantum correlations} between the two by comparing them. The presence or absence of a target, as well as its precise location and speed, can be inferred from these correlations \cite{Ref10}-\cite{Ref15}. The mathematical framework for this procedure is a \textbf{binary quantum hypothesis testing problem}. The detection process is modeled as a binary quantum hypothesis testing problem, so this becomes a problem of quantum distinguishability; thus, we use some techniques to measure quantum correlation, which I have listed below. 

The radar receiver distinguishes between two joint quantum states of the return and idler modes:

\begin{itemize}
  \item \( \rho_0 \): idler + noise (no target, uncorrelated)
  \item \( \rho_1 \): idler + return (target present, weak correlation)
\end{itemize}

The optimal measurement that minimizes the probability of error is given by the Helstrom bound,which can even be represented using the state:
\[
P_e = \frac{1}{2} \left( 1 - \frac{1}{2} \lVert \rho_1 - \rho_0 \rVert_1 \right)
\]

The \textbf{quantum fidelity} between the two states is:
\[
F(\rho_0, \rho_1) = \left( \mathrm{Tr} \sqrt{ \sqrt{\rho_0} \rho_1 \sqrt{\rho_0} } \right)^2
\]

The distance \textbf{ from the trace} is:
\[
D(\rho_0, \rho_1) = \frac{1}{2} \lVert \rho_0 - \rho_1 \rVert_1
\]

These metrics quantify the distinguishability of the quantum states, which guides us if the target is present or not, and thus data can be extracted from a quantum radar using Positive Operator-Valued Measures (POVMs) or unsharp measurements.

\bigskip

The expected power output of quantum radar systems, measured in decibels (dB): Quantum radar typically operates with extremely low power, around -130 to -100 dBm, due to single- or few-photon transmissions. To understand this, we use the following formula to convert from watts to decibels (dBm):
\[
\text{Power (dB)} = 10 \cdot \log_{10} \left( \frac{P}{P_0} \right)
\]
where \( P \) is the power in watts, and \( P_0 \) is the reference power (1 mW or \( 1 \times 10^{-3} \) watts).

For example, a power level of \( 10^{-13} \) watts can be converted to dB as:
\[
\text{Power (dB)} = 10 \cdot \log_{10} \left( \frac{10^{-13}}{10^{-3}} \right)
\]
\[
\text{Power (dB)} = 10 \cdot \log_{10} \left( 10^{-10} \right) = -100 \, \text{dBm}
\]
This corresponds to one possible power level in quantum radar systems, the range is around \textbf{-130 to -82 dBm}.

\bigskip
The power output compare to traditional radar systems: Traditional radars emit signals in the range of \textbf{+22 to +65 dBm}. Quantum radar’s low power reduces detectability and enhances stealth.

\bigskip

\begin{table}[]

\begin{tabular}{|l|l|}
\hline
\textbf{Prototype} & \textbf{Size (cm)} \\
\hline
IST Austria Microwave Quantum Radar & $\sim$100 \\
\hline
Munich Quantum Instruments Q one & 48.3 (19-inch rack) \\
\hline
Canadian DRDC Lab Prototype & $\sim$100--200 \\
\hline
Chinese CETC Mockup & Not specified \\
\hline
\end{tabular}
\label{tab:quantum_radar_dimensions}
\end{table}

\bigskip

The design constraints that affect the size of the radar:

 \begin{itemize}
    \item \textbf{Coherence}: Quantum states must be shielded from external disturbances. This requires stable optical setups and shielding to protect the system from thermal and electromagnetic fluctuations, which directly impacts the overall size of the system.
    \item \textbf{Cooling}: Many quantum radar systems rely on superconducting detectors that need to operate at extremely low temperatures. The cryogenic cooling systems required to achieve such low temperatures contribute significantly to the size of the radar system.
    \item \textbf{Optical Alignment}: Precise alignment of optical components, such as beam splitters and lenses, is essential for the effective operation of quantum radar. Misalignment can lead to signal losses or reduced performance, thus requiring more space for stable mounts and optical alignment systems.
    \item \textbf{Isolation}: Quantum radar systems must be isolated from external noise sources to maintain the integrity of the quantum states. This requires the use of Faraday cages, vibration damping, and electromagnetic shielding, all of which add to the physical dimensions of the system.
\end{itemize}

\bigskip

During the deployment process, it is necessary to handle the electromagnetic interference (EMI) and electromagnetic compatibility (EMC) issues in order to guarantee that the system will function effectively and without interruption. Through the application of appropriate shielding and grounding procedures, it is possible to alleviate these problems and guarantee compliance with regulatory standards. By carefully designing the layout of components and cables, as well as utilizing filters and surge protectors, engineers can minimize the impact of EMI and EMC on the system. Quantum radar systems are vulnerable to \textbf{Electromagnetic Interference (EMI)} due to their low-power operation. Weak quantum signals, such as single-photon or few-photon signals, are highly sensitive to external interference from classical electronics or RF systems \cite{Ref16}-\cite{Ref30}. To mitigate these effects, the following techniques are essential:

\subsection*{1. Shielding}
Shielding, using conductive materials like copper or aluminum, prevents external RF signals from interfering with quantum radar components. The effectiveness of shielding is quantified by \textbf{Shielding Effectiveness (SE)}, which depends on the thickness of the shielding and the wavelength of the interference:
\[
SE = 20 \log \left(\frac{d}{\lambda}\right)
\]
where \(d\) is the shielding thickness, and \(\lambda\) is the wavelength of the external interference.

\subsection*{2. Isolation}
Isolation involves using physical barriers such as Faraday cages and grounding techniques to reduce external noise. The \textbf{Isolation Factor (I)} is given by:
\[
I = \frac{N_{\text{ext}}}{N_{\text{isolated}}}
\]
where \(N_{\text{ext}}\) is the noise level before isolation and \(N_{\text{isolated}}\) is the noise level after isolation.

\subsection*{3. Narrowband Filtering}
Narrowband filtering allows only the desired frequencies to pass through, rejecting unwanted noise. The effectiveness of the filter is characterized by \textbf{Stopband Attenuation (SA)}:
\[
SA = 20 \log \left( \frac{A_{\text{stop}}}{A_{\text{pass}}} \right)
\]
where \(A_{\text{stop}}\) and \(A_{\text{pass}}\) are the amplitudes of the signals outside and inside the filter's passband, respectively.

\bigskip
 Improving signal processing algorithms and using more entangled photons could help fix the problems with quantum radar systems, along with the other ideas already discussed. Scientists are also looking into new technologies and materials that might make quantum radar systems more sensitive and effective.
 
\begin{itemize}
    \item \textbf{Shielding:} Employing conductive materials to block external electromagnetic fields.
    \item \textbf{Stinespring dilation theorem} provides a mathematical framework to model quantum channels as unitary operations on extended Hilbert spaces. In quantum radar systems, this theorem aids in understanding how environmental interactions and noise influence system performance. By leveraging this understanding, more effective EMI and EMC mitigation strategies, such as optimized shielding and filtering, can be developed.

    \item \textbf{Isolation:} Using Faraday cages and proper grounding to minimize external noise.
    \item \textbf{Filtering:} Implementing narrowband filters to allow only desired frequency bands.
    \item \textbf{Active EMI Cancellation:} Utilizing feedback and feedforward controlled filters to dynamically detect and neutralize interference.
    \item \textbf{Advanced Signal Processing:} Applying sophisticated algorithms to distinguish between genuine quantum signals and noise.
\end{itemize}

\bigskip

There are safeguards in place to reduce the likelihood of harm to humans from exposure to quantum radar and to guarantee that the technology is safe for use. To safeguard persons from potential danger, these procedures incorporate stringent power limitations, shielding, and radiation level monitoring.

\begin{itemize}
    \item \textbf{Directional Emission and Beam Control}: Quantum radar systems emit highly directional beams, focusing energy on specific targets while minimizing stray radiation. This design significantly reduces the risk of unintended human exposure.
    \item \textbf{Controlled Environments and Usage Protocols}: Quantum radar systems are primarily operated in controlled environments, such as research laboratories or military installations, where strict safety protocols are enforced to limit human exposure.
    \item \textbf{Thermal Effects Consideration
}: While quantum radars operate at low power levels, it's essential to consider thermal effects from prolonged exposure to electromagnetic fields. Studies have investigated both thermal and non-thermal biological effects of radar exposure.
    \item \textbf{ Redundancy and Fail-Safe Systems
}: To enhance safety, quantum radar systems incorporate redundancy and fail-safe mechanisms to prevent unintended emissions or system failures that could lead to unsafe exposure levels.
\end{itemize}

 The level of sensitivity can be expected from quantum radar technology:

\begin{itemize}
    \item \textbf{Quantum Radar Sensitivity:} Quantum radar can detect signals as weak as:
    \[
    P_{\text{min}} = 10^{-16} \, \text{W}
    \]
    which corresponds to detecting individual microwave photons under ideal conditions.
    
    \item \textbf{Sensitivity in Radar:} The minimum detectable signal power is determined by the system's noise level. Quantum radar exploits quantum effects like \textit{entanglement} and \textit{quantum correlations} to go below classical noise limits (e.g., thermal and shot noise).
    
    \item \textbf{Photon Energy and Count:}
    \begin{itemize}
        \item Photon energy at microwave frequency (\( f = 10 \, \text{GHz} \)):
        \[
        E_{\text{photon}} = h f = (6.63 \times 10^{-34}) \times (10^{10}) = 6.63 \times 10^{-24} \, \text{J}
        \]
        
        \item Number of photons detectable per second at \( P = 10^{-16} \, \text{W} \):
        \[
        N = \frac{P}{E_{\text{photon}}} = \frac{10^{-16}}{6.63 \times 10^{-24}} \approx 1.5 \times 10^7 \, \text{photons/sec}
        \]
    \end{itemize}
    This shows that quantum radar detects tens of millions of photons per second, far lower than classical radar, which detects trillions or quadrillions of photons per second.

\end{itemize}

\bigskip

The sensitivity impact target detection and identification:
\begin{itemize}
  \item \textbf{Higher sensitivity} means lower minimum detectable power ($P_{\text{min}}$).
  \item Improves \textbf{detection range}:
  \[
  R \propto \left( \frac{1}{P_{\text{min}}} \right)^{1/4} \Rightarrow 100\times \text{ better sensitivity} \Rightarrow 3.16\times \text{ range}
  \]
  \item Allows detection of \textbf{low-RCS targets} (stealth jets, small drones).
  \item Works in \textbf{noisy or jammed environments} using quantum correlations.
  \item Enhances \textbf{target identification}:
  \begin{itemize}
    \item Higher resolution
    \item Better Doppler detection
    \item Quantum fingerprinting of target features
  \end{itemize}
\end{itemize}

\bigskip

The size and weight of a typical quantum radar system:
\begin{table}[h!]
\centering

\begin{tabular}{@{}lll@{}}
\toprule
\textbf{Radar Type} & \textbf{Weight Range} & \textbf{Dimensions (Typical)} \\ \midrule

\textbf{Quantum Radar (Lab-scale)} & 20–200 kg & Large bench systems \\

\textbf{Quantum Radar (Industrial prototype)} & 50–100+ kg & Medium mobile units \\

\textbf{Quantum Radar (Compact/commercial)} & 2–10 kg & $\sim$30×20×13 cm \\

\textbf{Classical Radar (Airborne, e.g. STARLite)} & 20–65 kg & $\sim$1.2 ft\textsuperscript{3} \\

\textbf{Classical Radar (Ground-based, e.g. GM400)} & $<$10,000 kg & 20-ft ISO container \\

\bottomrule
\end{tabular}
\end{table}

\bigskip

Concept of Entanglement in Signal and Idler Photons: A signal photon (S) and an idler photon (I), which are entangled. These photons are generated simultaneously in such a way that the quantum state of the pair is described collectively, not individually.

The quantum state of this entangled pair is:
$$
|\psi_{SI}\rangle = \frac{1}{\sqrt{2}} \left( |0\rangle_S |0\rangle_I + |1\rangle_S |1\rangle_I \right)
$$
This state implies that:

1. Neither photon has a definite state individually.
2. When one photon (say, S) is measured and found in state \( |0\rangle \), the other (I) is
    immediately known to be in \( |0\rangle \), regardless of distance. Likewise, if S is measured in state \( |1\rangle \), then I is also in state \( |1\rangle \). This is a nonlocal quantum correlation—no classical system can reproduce this behavior.

Key Insight: Measurement on the signal photon instantly collapses the joint entangled state, revealing the state of the idler photon. This property is harnessed in quantum-enhanced technologies like quantum radar, where the idler is kept and compared against the signal to extract subtle target information.

Stepwise Operation

Step 1: Entangled Photon Generation:

A pair of entangled photons (signal S and idler I) is created using quantum dots in waveguides. The initial Bell-type state is: 
$$
|\psi_{SI}\rangle = \frac{1}{\sqrt{2}} \left( |0\rangle_S |0\rangle_I + |1\rangle_S |1\rangle_I \right)
$$
This state has perfect quantum correlations and is globally sensitive to phase shifts applied to only one photon.

Step 2: Signal Transmission and Target Interaction

The signal photon is transmitted toward a possible target. If it reflects off a target, it picks up a phase:
$$
|1\rangle_S \rightarrow e^{i\phi}|1\rangle_S
$$
Thus, the total state becomes:
$$
|\psi'_{SI}\rangle = \frac{1}{\sqrt{2}} |0\rangle_S |0\rangle_I + e^{i\phi}|1\rangle_S |1\rangle_I
$$
Even though the phase is applied only to the signal, the \textbf{global entangled state evolves}, making it detectable.

Step 3: Idler Retention and Phase-Based Inference

The idler photon is retained in the lab. After the signal photon interacts with the environment and returns, the entangled state remains preserved. However, the interaction may induce a relative phase shift in the signal mode. This relative phase alters the global entangled state, even though local entanglement is maintained. As a result, joint measurement of the idler and signal enables inference of this phase shift.

Step 4: Quantum Hypothesis Testing
We apply binary quantum hypothesis testing:
\begin{itemize}
    \item H0: No target present (only noise) state is separable: \( \rho_0 = \rho_I \otimes \rho_{noise} \)
    \item H1: Target reflected signal state is entangled with a phase-modified signal:
    $$
    \rho_1 = |\psi'_{SI}\rangle\langle\psi'_{SI}|
    $$
\end{itemize}
The radar performs a joint measurement to determine which hypothesis is true.

Step 5: Final Inference and Target Detection
If the fidelity is low, it may be due to noise; however, a phase change caused by a
target does not destroy entanglement, so the fidelity remains higher compared to
classical noise:
$$
F(\rho_0, \rho_1) = \text{Tr} \sqrt{\rho_0 \rho_1} \sqrt{\rho_0}^2
$$

Core Functions in Quantum Radar: Phase-Based Inference

Quantum radar uses entangled states to detect phase-based differences between target returns and background noise. This enables three key quantum capabilities:

• Stealth Detection: Targets induce coherent phase shifts in one arm of the entangled state, e.g.,
$$
|\Psi_\phi\rangle = \frac{1}{\sqrt{2}} \left( |00\rangle + e^{i\phi}|11\rangle \right)
$$
These shifts preserve entanglement and reduce fidelity with the no-target state, allowing detection even with weak reflections.

• Signal Noise Rejection: Random environmental photons or thermal noise produce incoherent, uncorrelated states. These destroy off-diagonal coherence and do not match the expected quantum signature, so they are filtered out during photon comparison.

• Environmental Compensation: Predictable phase shifts from atmospheric effects are tracked using the idler photon. Since the idler retains the original quantum state, deviations due to environment can be compensated, isolating the target’s effect.

Unlike classical radar, quantum radar performs inference via entangled phase analysis, enabling robust detection under stealth and noisy conditions.

The factors influence deployment logistics:

\begin{itemize}
  \item \textbf{Heavy lab-scale systems} (20-200 kg) are difficult to transport and require stable, vibration-free environments — limiting them to fixed or lab-based setups.
  \item \textbf{Cryogenic cooling needs} add bulk and complexity, making airborne or satellite deployment challenging without compact cooling advances.
  \item \textbf{Industrial prototypes} (50-100+ kg) can be vehicle-mounted, but require shielding from environmental noise and motion.
  \item \textbf{Compact systems} (2-10 kg) are better suited for mobile and tactical use but often trade off sensitivity or range.
  \item \textbf{Classical radars} are generally easier to deploy on diverse platforms due to lack of quantum-specific constraints like entanglement preservation and cooling.
\end{itemize}

\bigskip

The noise calculated in quantum radar systems:

\begin{itemize}
  \item Noise in quantum radar systems is primarily influenced by thermal background radiation, photon statistics, and quantum noise.
  \item \textbf{Thermal Background Noise:} 
    \[
    \bar{n} = \frac{1}{e^{\frac{h \nu}{k_B T}} - 1}
    \]
    where \( h \) is Planck's constant, \( \nu \) is the frequency, and \( k_B \) is the Boltzmann constant. This represents the average photon number due to ambient thermal radiation.
  
  \item \textbf{Photon Statistics:} Quantum radar systems often employ squeezed states or entangled photon pairs to reduce noise and improve SNR, surpassing the standard quantum limit.
  
  \item \textbf{Signal-to-Noise Ratio (SNR):}
    \[
    \text{SNR} = \frac{P_{\text{signal}}}{P_{\text{noise}}}
    \]
    This equation quantifies the ratio of signal power to noise power in the radar system.

  \item \textbf{Example Calculation:} 
    \[
    \text{SNR} = \frac{10^{-16}}{10^{-15}} = 0.1
    \]
    An SNR of 0.1 implies that the signal is weaker than the nose.
\end{itemize}

\bigskip

The techniques are used to tackle noise and effectively identify objects

 \begin{itemize}
  \item \textbf{Homodyne Detection:} A technique that measures the phase of a signal by comparing it to a reference signal, allowing detection of weak signals with high precision.
  \item \textbf{Quantum Illumination:} Uses entangled photons to illuminate a target and improve detection in noisy environments by exploiting quantum correlations between signal and noise.
  \item \textbf{Entropy Filtering:} A method of filtering signals based on entropy to remove noise by isolating the relevant information that carries the target's signature.
  \item \textbf{Time-Gating:} A technique where the radar system only collects data during a specific time window, effectively reducing background noise and increasing the signal-to-noise ratio.
  \item \textbf{Stinespring Dilation:} A mathematical technique used to represent quantum operations, enabling the optimal processing of noisy quantum signals by transforming them into a higher-dimensional space, improving noise handling.
\end{itemize}

\bigskip

The quantum radar differs fundamentally from classical radar systems:

\begin{itemize}
  \item \textbf{Entanglement:} Quantum radar utilizes entangled photon pairs to enhance detection sensitivity, especially in noisy environments. This allows the radar to detect weak signals that would be overwhelmed by noise in classical radar systems.
  \item \textbf{Improved Noise Resilience:} Quantum radar can leverage quantum properties, like quantum squeezing or quantum illumination, to improve signal-to-noise ratios (SNR), even in high-noise conditions, making it highly effective in cluttered or jamming-prone environments.
  \item \textbf{Enhanced Range and Resolution:} The use of quantum states allows for better range detection and finer resolution, even at lower power levels, which classical radar systems struggle to achieve under similar conditions.
\end{itemize}

\end{document}